\documentclass{article}

\usepackage{arxiv}
\usepackage{hyperref}
\usepackage[utf8]{inputenc} % allow utf-8 input
\usepackage[T1]{fontenc}    % use 8-bit T1 fonts
\usepackage{hyperref}       % hyperlinks
\usepackage{url}            % simple URL typesetting
\usepackage{booktabs}       % professional-quality tables
\usepackage{amsfonts}       % blackboard math symbols
\usepackage{nicefrac}       % compact symbols for 1/2, etc.
\usepackage{microtype}      % microtypography
\usepackage{graphicx}
\usepackage{tabularx}
\usepackage{multirow}
\usepackage{parskip}
\usepackage{subcaption}
\usepackage{graphicx}
\usepackage{enumitem}
\usepackage{cite}
\title{TIDF-DLPM: Term and Inverse Document Frequency based Data Leakage Prevention Model}
\date{}   

\author{ \href{https://orcid.org/0000-0003-3746-6034}{\includegraphics[scale=0.06]{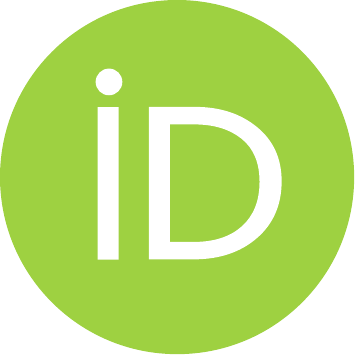}\hspace{1mm}Ishu Gupta*}
	\\
	Cloud Computing Research Center,
	Department of Computer Science and Engineering\\ 
	National Sun Yat-sen University
	Kaohsiung, Taiwan\\
	\texttt{ishugupta23@gmail.com} \\
	\And Sloni Mittal \\
	Department of Computer Applications, National Institute of Technology,
	Kurukshetra, India, 136119\\
	\texttt{mittalsloni09@gmail.com}
	\And Ankit Tiwari \\
	Department of Computer Applications, National Institute of Technology,
	Kurukshetra, India, 136119\\
	\texttt{ankitme8694@gmail.com}
	\And Priya Agarwal \\
	Department of Computer Applications, National Institute of Technology,
	Kurukshetra, India, 136119\\
	\texttt{priyaagarwal928@gmail.com}
	\And 	\href{https://orcid.org/0000-0002-8053-5050}{\includegraphics[scale=0.06]{orcid.pdf}\hspace{1mm}Ashutosh Kumar Singh} \\
	Department of Computer Applications,
	National Institute of Technology,
	Kurukshetra, India 
	136119\\
	\texttt{ashutosh@nitkkr.ac.in}
}

\hypersetup{
pdftitle={A template for the arxiv style},
pdfsubject={q-bio.NC, q-bio.QM},
pdfauthor={David S.~Hippocampus, Elias D.~Striatum},
pdfkeywords={First keyword, Second keyword, More},
}

\begin{document}

\maketitle
\begin{abstract} 
	 Confidentiality of the data is being endangered as it has been categorized into false categories which might get leaked to an unauthorized party. For this reason, various organizations are mainly implementing data leakage prevention systems (DLPs). Firewalls and intrusion detection systems are being outdated versions of security mechanisms. The data which are being used, in sending state or are rest are being monitored by DLPs. The confidential data is prevented with the help of neighboring contexts and contents of DLPs. In this paper, a semantic-based approach is used to classify data based on the statistical data leakage prevention model. To detect involved private data, statistical analysis is being used to contribute secure mechanisms in the environment of data leakage. The favored Frequency-Inverse Document Frequency (TF-IDF) is the facts and details recapture function to arrange documents under particular topics. The results showcase that a similar statistical DLP approach could appropriately classify documents in case of extent alteration as well as interchanged documents.
	\end{abstract}
%https://www.overleaf.com/project/6129194805bd1181f52dde18

% keywords can be removed

\section{Introduction}
Data security ensures Confidentiality, Authenticity, and integrity for the documents sent across the network which further arises the availability and classification issues for the documents which compromise the security to a further extent \cite{Saxena,Arora}. This can further lead to trade secrets, bank details, the privacy of patients, health records, security of accounts and the list goes on due to inappropriate classification of documents in categories different from the one they belong which violates the confidentiality of the users \cite{DT-ILIS,Gautam}. The disclosure is expanding with respect to dimensions and action has been stated in the recent report \cite{Arthur,CSA}. For example, eBay one of the biggest online shopping websites experienced one of the major setback leaks in history when more than 145 million customers' personal details such as phone numbers, names, email ids, etc. were theft \cite{Kaur2017,IOSR}. As an outcome, a massive reset account password was carried out from the customer’s end \cite{GUIM-SMD,Holistic}.

To eliminate these problems executants and scholars have illuminated techniques and methodology to safeguard confidential data which is mainly acknowledged by the term Data Leakage Prevention Systems (DLPSs) \cite{IDS,Preetesh}. DLPSs mark the continuous investigating of confidential data which lacked standard security mechanisms such as Intrusion detection and firewalls \cite{OnILIS,ICCNSJapan}. The confidential data is endlessly supervised by DLPSs irrespective of data \lq\lq in transit\rq\rq or \lq\lq in use\rq\rq \cite{Jalwa,CC,Harsh}. The analysis of intimate data with respect to various contentment and context is done by DLPSs for such kind of detection. Studying components such as sender, receiver, format, time, and data size fall under contextual analysis \cite{IJNSA, Rajat}. The regular expression, fingerprinting, and statistical analysis of content are under content analysis \cite{Jadon,EPS}. The perception is made when duplicate or contents of the intimate data is being retrieved, broadcasted, or pre-owned without the consent of the user \cite{Ayushi,Godha}. The drawback of the regular expression is that it has restricted extend so only rule-based items are in the ease of satisfaction \cite{SELI,Hybrid}. The major drawback of fingerprinting is the vulnerability of being different from the one suspected \cite{JISE,Sharma}.

The use of DLPs for curative actions such as caution, chunks, encryptions, and audits for safeguarding these tasks \cite{Tiwari,HISA-SMFM}. Identification takes place when duplicates or parts of the intimate data are monitored, retrieved, or transferred without validation. Statistical analysis, regular expressions, and fingerprinting are used to recognize duplicates of confidential data \cite{JCOMSS,Kamal}. Advanced Fingerprinting is the fingerprints produced by the data hashing have some faults to overcome this we use modified full data fingerprinting \cite{PCS,Animesh}. In usual fingerprints the major culpability is that they are unsafe and can be dodged with smaller changes to the authentic one: that’s why to generate altered fingerprints k-skip-n-gram come into scenario \cite{singh2020survey}. Even after the altercation k-skip-n-gram is the vigorous way for detecting the authentic data (like subtraction, addition, synonyms) \cite{IJAST,Kaur2018}. In this method, both intimidated and non-intimidated docs are managed to create fingerprints, in which intimidated docs create non-intimidated k-skip n-grams that will help in removing unwanted n-gram in the confidential documents \cite{Kesharwani,Khushbu}. In almost all scenarios this suggested procedure outperformed the nominal fingerprinting method. For all confidential and non-confidential comprehensive indexing is required. Additional storage is a major drawback, as well as processing competence, are needed \cite{Confidentiality,BatraGarima}. To search acceptable, resemble documents Binary codes as a memory address (BCMA) is used and was able to get resembling documents in an insignificant fraction of time taken by LSH \cite{Sloni,Nishad}. The main limitation in this is that it is only capable of getting documents based on a specific topic like government borrowing and disaster \cite{MACI,Vartika}. 
This paper deals with and scrutinizes the accomplishment of using statistical analysis in identifying intimate data semantics. Term Frequency-Inverse Document Frequency (TF-IDF) is the term weighting function on which the DLP model is based. 

\section{Related Work}
For identifying data semantics, there are tremendous approaches guided with the help of promoted essence of fingerprinting and statistical analysis; for example, advanced fingerprinting \cite{Lewis,Wakefield,MLPAM,kaur2017comparative}. Advanced Fingerprinting is the fingerprints produced by the data hashing have some faults to overcome this we use modified full data fingerprinting. Nominal fingerprints' main drawback is that they are unsafe and can be dodged with smaller changes to the ordinary one: that’s why k-skip-n-gram comes into the picture to generate fingerprints that are altered. The k-skip-n-gram initiates a vigorous way for detecting the nominal data even after alternation (like subtraction, addition, synonyms). In this method, both privileged and non-privileged documents are managed to create fingerprints, in which non-privileged documents create non-privileged k-skip-n-grams that will help in removing unwanted n-gram in the intimidating documents.
In almost all scenarios ordinary full fingerprinting methods are outperformed. For all intimidated and non-intimidated comprehensive indexing is required. The major drawback for this is that additional storage and dealing potentiality are needed.  

Another method was using a Support Vector Machine (SVM) which uses an SVM algorithm are used to categorize three types of data: enterprise private, enterprise public, and non-enterprise. The percentage of data leakage is 97\% with 3\% false negative. But this method only identifies private and the public failed to detect more flexible classification levels like top classified, classified, and privileged. Thus, this method will no longer be in use and make the enforcement of security policies difficult \cite{Cohen}. 
TrendMicro Locality Sensitive Hashing (TLSH) was introduced to make one change in data detection accuracy, quartiles method is come into play to pathway the counting bucket and has a sliding window of 5 bytes. TLSH is tested against insertion, deletion, edition. It has also performed text modification identification to obtain higher precision and recall that LSH was presented \cite{Matic}.
 Next approach Binary codes were used as a memory address (BCMA) is used to search resembling documents and was successful to obtain resembling documents in a negligible fraction of time taken by LSH. The main restriction in this is that it is only capable of getting documents based on specific topics like disaster and government borrowing. Challenge is trying to get sub-topics within a general topic \cite{Morgan}.

\section{Data Leakage Prevention Model}
The deployment model for data leakage prevention which showcases the various data states, analyzes the various DLP’s context and further various remedial actions which have been taken; is represented in diagrammatic form in Fig. 1. DLP model which is based on semantic approach, to classify data a statistical data leakage prevention (DLP) model is developed. The objective behind this DLPs technique work is to use neighboring context and content of personal private data to identify and secure malicious access to confidential data. Despite the data is being used by the intended users or while data is transferred internally or externally between communication channels. Data leakage is defined as a strategy that guarantees sensitive and confidential data will not go outside the organization. To detect confidential data semantic, a statistical data analysis model is developed for DLP. The most-popular term weighting function TF-IDF is employed to measure tested document and the aggregate statistics contained in it. The predefined categories cluster the related topic together which is basically considered to be the main aim of clustering. Considering that every grouping has a confidential level, documents with restricted secrecy levels can be easily detected and actions will take like chunks, observant, and quarantines. The document can be easily identified if a user is attempting malicious access, use, and alter it with the use of fingerprinting and regular expression. However, the user alters the document by inserting, deleting, or exchanging words, lines, paragraphs. Even in the case of robust data fingerprinting, identifying documents will be considered to be a challenging event statistical analysis can be used to calculate and approximate the private data with the help of the DLP model. 

\begin{figure}[h]
	\centering
	\includegraphics[width=0.55\textwidth]{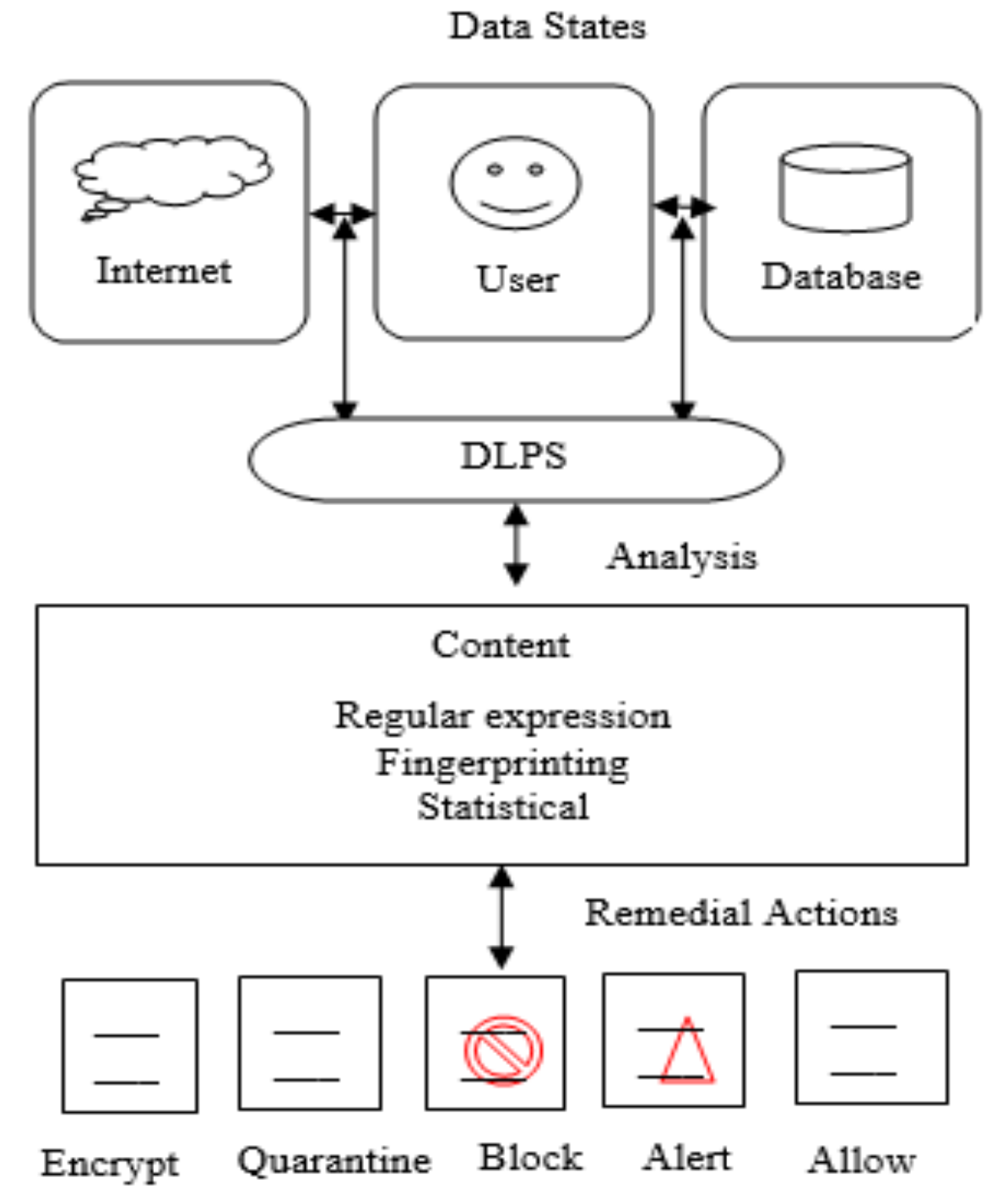}
	\caption{Deployment data leakage avoidance model}
	\label{fig:my_label1}
\end{figure}

\subsection{Centroid based Document Classifier}
In the centroid-based document model, the documents are showcased with the help of vector space. In this model, every text in the document is assumed to be the vector in term-space. In a simple way, every document is presented by the Term Frequency (TF). Term frequency can be defined as it is used in finding connection with information retrieval and display occurring of the term frequently in a document how frequently. TF shows the count of particular words within the whole document. This value is often mentioned in Inverse document frequency (IDF). This is basically used for data mining and information retrieval.

\subsection{Term Frequency}
The number of occurrences of a particular term $t$ in a document $d$ is represented by the term frequency.
If a particular word repeats many times in a document it becomes more important logically. For this, we use a vector to represent a bag of words. Model and ordering of terms are not compulsory.
For each value term, there is an entry corresponding frequency will be given. The presentation of documents as vectors in a common vector space is known as a space model.
The classification representation along the centroid in the document has been showcased in Fig. 2 for better visualization and understanding.

\begin{figure}[h]
   \centering
    \includegraphics[width=0.45\textwidth]{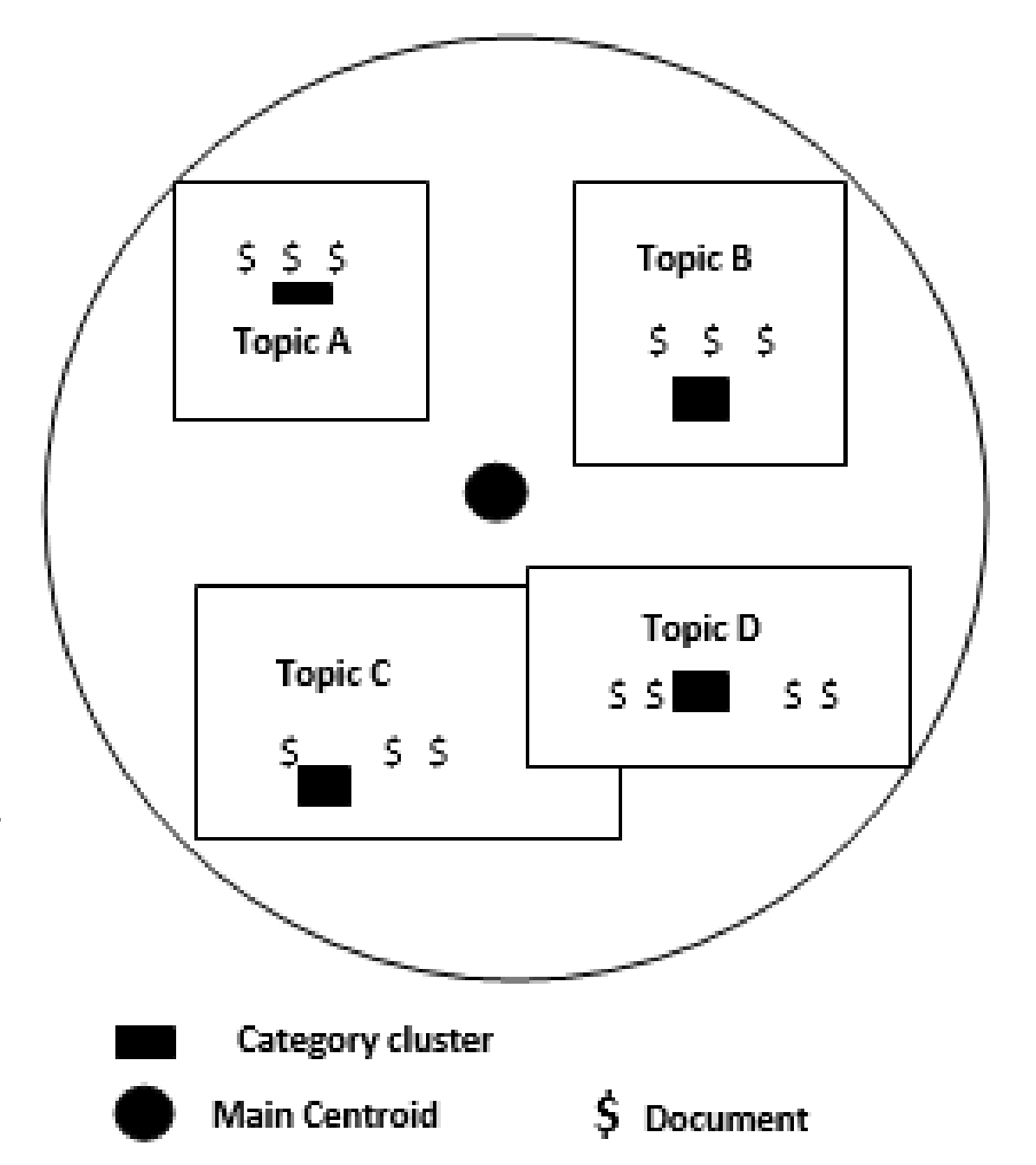}
    \caption{Classification representation}
    \label{fig:my_label2}
\end{figure}

\subsection{Inverse Document Frequency}
 This plays an important role in measuring how important a term is. The main focus of searching is to find some suitable documents for the query. Till now we know about TF which considers all the terms equally important that's why we are not using TF here to calculate the weight of a term in a document. As we know common words appear more in the document but they have less importance. Hence, we have to weigh down the important terms while considering the rare ones. So inverse document frequency is defined as the number of documents in the corpus divided by the document frequency of a term. In this paper, we will be talking about both term frequency and inverse document frequency commonly known as TIDF, and combine them to get an ultimate score of term $t$ in document $d$.  
 
\section{Experiments}
Centroid-based classifier interpretation has been assessed by contrasting it against numerous numbers of document gathering techniques such as k-nearest-neighbor, naive Bayesian, and C4.5. TIDF vector-space documents showcasing has been used in the case of k-nearest-neighbor. Multinomial event model commonly known as Rainbow is used for Bayesian results formulation. A confined altered sort of C4.5 algorithm having the ability to manage sparse data sets has been used for gathering the outcome of C4.5.

\subsection{Document Gatherings}
The comprehensive features of numerous documents gathered for the experiments are presented here and it is to be noted that to remove similar words, we use a stop-list.
The data sets for badminton, soccer, cricket, tennis are gathered and collected from different websites articles which were converted into text format for making them eligible for data sets. The website used for gathering these is badminton-information.com, cricbuzz.com, tennis-online, and eurosoccerfan.com. These are the various web page that corresponds to the data set collections used in our experimental results.

\subsection{Classification Performance}
Since document data sets have been formulated where single class label has been presented in each document and true categorization demonstration has been performed by the means of classification accuracy.
Our formulation shows the categorization of various algorithms on contrasting data sets. The average precision outcome formulation of 10 experiments has been showcased in our paper. The demonstration of the documents which is known to be 80\% are haphazardly selected as the training set and the remaining 20\% as the test set. k-nearest neighbor schemes, C4.5, naive Bayesian have been showcased in the starting three rows of Table 1 and the fourth row contains the centroid-based categorization. It can be seen from Table 1 which categorizes the accuracy of classifiers that the centroid based categorization outperforms the Naïve Bayes, C4.5 and k-Nearest Neighbor for the data sets i.e. Cricket (Crick), Badminton (Badm), Tennis (Tenn.), and Soccer (Socc) collected from websites. 

\begin{table}[h]
    \centering
    \caption{Categorization Accuracy of Classifiers }
    \begin{tabular}{cc|c|c|c|}
        \hline  
        \multicolumn{1}{|c|}{\multirow{2}{*}{\textbf{Algorithm}}}  & \multicolumn{4}{c|}{\textbf{Categories}} \\ 
        \cline{2-5}
        \multicolumn{1}{|c|}{} & \textbf{Crick} & \textbf{Tenn} & \textbf{Badm} & \textbf{Socc}\\ 
        \hline
        \multicolumn{1}{|c|}{\begin{tabular}[c]{@{}c@{}}NaiveBayesian\end{tabular}} & 89.3 & 91.2 & 84.3 & 72.3 \\      \hline
        \multicolumn{1}{|c|}{\begin{tabular}[c]{@{}c@{}}k-Nearest N.\end{tabular}} & 85.8 & 87.5 & 77.5 & 84.6 \\ \hline
        \multicolumn{1}{|c|}{\begin{tabular}[c]{@{}c@{}}Centroid Based\end{tabular}} & 91.8 & 93.9 & 82.7 & 94.2 \\
        \hline\\
    \end{tabular}
    \label{tab:my_label1}
\end{table}

\section{Analysis}
\subsection{Classification Prototype}
The centroid-based classifier is unpredictably good. The aim of this section is to give an overview of how a Centroid-based classifier outperforms various algorithms and performs well in all cases. A centroid-based classifier regulates the resemblance between the specific class and a test document. Basically, the mean resemblance is being calculated between the test document and all the additional documents in the current class. If the magnification is higher which corresponds to a small level of resemblance between the documents whereas if the magnification is lower it corresponds to the high level of resemblance among the documents.

\subsection{Comparison Among Classifier}
Differentiation of various classification algorithms using some sample pairs tested values in Table 2 shows that the better performance is given by row classifier and performs worse than classifier represented in the column. The analytical notable results are summarized using the sample paired test, in which various classification algorithms are used. It shows that classifiers who are performing best, are worse than other classifiers. From Table 2, we got to know that centroid based is 3 times better than naïve Bayesian, worse in one data set. Alike centroid-based is 3 times better than c4.5, worse 1 times in data sets. So what we are doing here is that better performance is given by row which is being worse than classifier in the column.

\begin{table}[h]
	\centering
	\caption{Performance Comparison}
	\begin{tabular}{|c|c|c|c|}
		\hline
		& \textbf{NB}  &  \textbf{KNN}  &  \textbf{C4.5}  \\
		\hline
		Centroid Based  &   3/1 &  4/1  &  3/1 \\
		\hline
		KNN &   & 3/1 & 3/1 \\
		\hline
		Naive Bayesian &  &  & 4/1 \\
		\hline
	\end{tabular}
	
	\label{tab:my_label2}
\end{table}

\section{Conclusion}
As malicious activities during transmission over the network are increasing rapidly, there is a need for additional security to the message sent over the network given by categorizing the data. This paper provides a review of some common classifiers that are used to classify documents into the correct category so that confidentiality is not compromised. Data leakage can happen because of using a bad clustering approach. So, to prevent this we used an approach named centroid document classifier in which we cluster data properly that helps in data leakage prevention. centroid-based document classification is steady and maintainable, it outperforms other classification algorithms on various data set. 

\end{document}